\documentstyle[preprint,aps,psfig,cite,tighten,amsfonts]{revtex}

\newcommand{\nc}{\newcommand}
\nc{\grad}{\nabla}  
\nc{\tr}{\mathop{\rm Tr}}
\nc{\half}{{1\over 2}}
\nc{\third}{{1\over 3}}
\nc{\be}{\begin{equation}}
\nc{\ee}{\end{equation}}
\nc{\bea}{\begin{eqnarray}}
\nc{\eea}{\end{eqnarray}}

\def\Tr{{\rm Tr}}
\nc{\dint}[2]{\int\limits_{#1}^{#2}}
\nc{\D}{\displaystyle}
\nc{\PDT}[1]{\frac{\partial #1}{\partial t}}
\nc{\tw}{\tilde{w}}
\nc{\tg}{\tilde{g}}
\nc{\newcaption}[1]{\centerline{\parbox{5.6in}{\caption{#1}}}}
\def\href#1#2{#2} 

\def\beq{\begin{eqnarray}}   \def\eeq{\end{eqnarray}}

\def\lsim{\mathrel{\rlap{\lower3pt\hbox{\hskip0pt$\sim$}}
    \raise1pt\hbox{$<$}}}         
\def\gsim{\mathrel{\rlap{\lower4pt\hbox{\hskip1pt$\sim$}}
    \raise1pt\hbox{$>$}}}         

\def\Z{{\mathbb{Z}}}
\def\C{{\mathbb{C}}}
\def\R{{\mathbb{R}}}
\def\Id{\hbox{1\kern-.23em{\rm l}}}

\nc{\al}{\alpha}
\nc{\ga}{\gamma}
\nc{\de}{\delta}
\nc{\ep}{\epsilon}
\nc{\ze}{\zeta}
\nc{\et}{\eta}
\renewcommand{\th}{\theta}
\nc{\Th}{\Theta}
\nc{\ka}{\kappa}
\nc{\la}{\lambda}
\nc{\rh}{\rho}
\nc{\si}{\sigma}
\nc{\ta}{\tau}
\nc{\up}{\upsilon}
\nc{\ph}{\phi}
\nc{\ch}{\chi}
\nc{\ps}{\psi}
\nc{\om}{\omega}
\nc{\Ga}{\Gamma}
\nc{\De}{\Delta}
\nc{\La}{\Lambda}
\nc{\Si}{\Sigma}
\nc{\Up}{\Upsilon}
\nc{\Ph}{\Phi}
\nc{\Ps}{\Psi}
\nc{\Om}{\Omega}
\nc{\ptl}{\partial}
\nc{\del}{\nabla}
\nc{\ov}{\overline}
\nc{\gsl}{\!\not}
\nc{\bi}[1]{\bibitem{#1}}
\nc{\fr}[2]{\frac{#1}{#2}}
\nc{\dsl}{\partial\!\!\!\!\!\!\not\,\,}
\nc{\gm}{\mbox{$\gamma_{\mu}$}}
\nc{\gn}{\mbox{$\gamma_{\nu}$}}
\nc{\Le}{\mbox{$\fr{1+\gamma_5}{2}$}}
\nc{\Ri}{\mbox{$\fr{1-\gamma_5}{2}$}}
\nc{\GD}{\mbox{$\tilde{G}$}}
\nc{\gf}{\mbox{$\gamma_{5}$}}
\nc{\Ima}{\mbox{Im}}
\nc{\Rea}{\mbox{Re}}

\nc{\av}{\langle \ph\rangle}
\nc{\ntwo}{${\cal N}\!\!=\!2\;$}
\nc{\none}{${\cal N}\!\!=\!1\;$}
\nc{\nfour}{${\cal N}\!\!=\!4\;$}

\def \bi{\bibitem}
\nc{\rf}[1]{(\ref{#1})}
\def \del{\partial}

\begin{document}
\draft
\preprint{
\vbox{\hbox{TPI-MINN-02/10}
      \hbox{UMN-TH-2050/02}
      \hbox{DAMTP-2002-44}
      \hbox{hep-th/0205083}}}

\title{Counting Domain Walls in \boldmath{${\cal N}\! =\!1$}
Super Yang-Mills Theory}

\author{
Adam Ritz$^{\,a}$, Mikhail Shifman$^{\,b}$, and
  Arkady Vainshtein$^{\,b}$
}
\address{
$^a$Department of Applied Mathematics and Theoretical
Physics,
  Centre for Mathematical Sciences, University of Cambridge,
Wilberforce Rd., Cambridge CB3 0WA, UK\\
$^b$Theoretical Physics Institute, University of Minnesota,
116 Church St SE,\\ Minneapolis, MN 55455, USA
}
\maketitle
\thispagestyle{empty}
\setcounter{page}{0}
\begin{abstract}
  
  We study the multiplicity of BPS domain walls in ${\cal N}\! =\!1$
  super Yang-Mills theory, by passing to a weakly coupled Higgs phase through
  the addition of fundamental matter. The number of domain walls 
  connecting two specified
  vacuum states is then determined via the Witten index of the 
  induced worldvolume theory, which is invariant under the deformation
  to the Higgs phase. The worldvolume theory is a sigma model 
  with a Grassmanian target space
  which arises as the coset associated with the global symmetries
  broken by the wall solution. Imposing a suitable infrared regulator,
  the result is found to agree with recent work
  of Acharya and Vafa in which the walls were realized as wrapped
  $D4$-branes in IIA string theory.

\end{abstract}

\newpage


\section{Introduction}
\label{intro}

It is well known that \none super Yang-Mills (SYM) theory with gauge
group $G$ exhibits $h$ distinct vacua where $h$, the dual Coxeter
number of the group, is equal to the adjoint Casimir invariant $T_G$. This
vacuum structure results from the spontaneous breakdown
$\Z_{2T_G}\rightarrow \Z_2$ of the discrete $\Z_{2T_G}$ remnant of the
anomalous U$(1)_R$ symmetry. As initially suggested by Witten
\cite{Witten1}, the relevant order parameter is the gluino condensate
$\langle\lambda^2\rangle \equiv \langle 0| \,{\tr}\,\lambda^2
|0\rangle\,$, first demonstrated to be nonzero within the framework of
the Veneziano-Yankielowicz effective Lagrangian \cite{Veneziano}.
Subsequently, the exact calculation of this condensate
\cite{ADS,nsvz,SVone} was performed via a controlled deformation to
weak coupling. The result takes the form
\begin{equation}
 \langle\lambda^2\rangle_k = \frac{16\pi^2}{g^2} M_{\rm PV}^3 
 \exp\left(\frac{2\pi i (\ta + k)}{T_G}\right) = 3 T_G\,
 \La^3\exp\left(\frac{2\pi i k}{T_G}\right)\,, \label{gcond}
\end{equation}
where $\ta=4\pi i/g^2 +\th/2\pi$ is the complex gauge coupling,
$M_{\rm PV}$ is the Pauli--Villars regulator scale, and 
the subscript $k=0,...,T_G-1$ marks the $k$-th vacuum. 
The second equality defines the renormalization 
group invariant dynamical scale $\La$.

The presence of discrete vacuum states implies the 
existence of topologically stable domain walls interpolating 
between them. Moreover, one expects BPS saturated domain walls to
exist since the \none supersymmetry algebra contains a 
central charge ${\cal Z}$ whose operator form is \cite{DSone,Stwo,CS}
\begin{equation}
{\cal Z}=\frac{T_G}{8\pi
^2}\int\! {\rm d} z \,\frac{\partial}{\partial z}\,
\lambda^2\,,
\label{symcc}
\end{equation}
where $z$ is the spatial coordinate perpendicular to the wall plane.
The expectation value of the central charge is nonvanishing
for states in 
the sector $|kn\rangle$, corresponding to domain 
wall configurations interpolating between the $n^{th}$ and $(k+n)^{th}$ vacua,
\begin{equation}
{\cal Z}_{k\,n}=\langle k\,n|\,{\cal Z}\,|k\,n\rangle=\frac{T_G}{8\pi
^2}\left(\langle\lambda^2\rangle_{k+n}-\langle\lambda^2\rangle_n\right)\,.
  \label{Zkn}
\end{equation}
The absolute value of ${\cal Z}_{k\,n}$ provides a
lower bound for the mass per unit area (i.e.\ the wall tension) 
of  $|kn\rangle$--sector states. This bound is 
saturated by BPS domain walls
whose tension $T$ (making use of (\ref{gcond}), (\ref{Zkn})) 
is given by 
\beq
T_k =|{\cal Z}_{k\,n}| 
  = \frac{3}{4\pi^2}\, T_G^2\,\La^3 \sin\frac{\pi k}{T_G}\;,
 \label{tension}
\eeq
which depends only on $k$. We will refer to  these BPS walls as
$k$-walls. They preserve two of the four supercharges, and thus form short
1/2--BPS multiplets containing one bosonic and one fermionic state.

In this paper, we address the problem of counting the number of such BPS 
supermultiplets for domain walls interpolating between two given vacua in \none
SYM theory, and we limit ourselves henceforth to gauge group SU($N$) where 
$T_{{\rm SU(}N{\rm )}}=N$. The number  of BPS multiplets $\nu_{k}$ is
counted by the CFIV index \cite{cfiv,fi,cv},
\be
 \nu_k = {\rm Tr}_{\,k\,n}\left[ F(-1)^F\right]\,, \label{cfivind}
\ee 
where $F$ is the fermion number, and the trace (suitably defined) 
runs over all states in the $|kn\rangle$--sector. The index counts
walls as solitonic objects in 3+1D, and thus there is an implicit
infrared regulator required to make the total wall mass finite.

An alternative way to view the CFIV index is that it counts the
number of supersymmetric vacua in the 2+1D theory induced on the 
worldvolume of the wall. We can then equivalently rewrite $\nu_k$ as
given by the Witten index \cite{Witten1} of the worldvolume theory,
\be
 \nu_k = {\rm Tr}_{\,\rm WV}\left[(-1)^F\right]\,. \label{wind}
\ee 
It is this alternative point of view which 
will be particularly useful for us.

The status of $\nu_{k}$ as an index refers to its invariance
under smooth $D$-term deformations, as shown in \cite{cfiv}, and it
is this feature which makes it a valuable quantity in theories with 
four supercharges, where $D$-terms are generally unconstrained. 
The index does, however, depend on the $F$-terms but these are
quantities over which \none supersymmetry exercises some control. 

For a strongly coupled system like \none SYM theory, the invariance of the
index under various deformations is crucial for the counting problem
to be tractable. It is clear that to make progress in evaluating 
(\ref{cfivind}), we should first deform the theory to a weak coupling 
regime where the $F$-terms upon which the index depends -- in this
case the holomorphic superpotential -- are calculable. Our strategy will
then be to determine $\nu_k$ in this regime, and then rely on
holomorphy of the superpotential, and the independence of
$\nu_k$ on the $D$-terms to return to pure SYM theory and deduce the 
corresponding wall spectrum. To this end, there are many 
possible deformations one could consider. Before outlining the 
particular route we will follow, it is appropriate to 
consider some other recent approaches to this problem, which 
in part motivated our work.\footnote{\,For additional related work 
on domain walls 
see \cite{walls1,kaplunovsky,wallsN,ASone,CM,CHMM,CHMM2,AStwo}.} 

One particular deformation, in which there has been considerable
recent interest \cite{vafadual,a1,amv,av,aw}, involves a geometric
realization of \none gauge theories as the low energy decoupling limit
of M-theory on a 7-manifold of $G_2$ holonomy.  Making use of a smooth
geometric transition in the moduli of the $G_2$-manifold \cite{amv,aw}
(see also \cite{civ,gt}) 
leads to a tractable large volume regime which exhibits many of the
features of the confining phase of SYM theory \cite{vafadual,a1}. For
SU($N$), the dual 7-manifold is topologically ${\R}^4\times S^3/\Z_N$,
and Acharya and Vafa \cite{av} proposed that in this context BPS walls
correspond to $M5$-branes wrapping the Lens space $S^3/{\Z}_N$.
Reducing to IIA string theory via the fibre of the Hopf map, leads to
$D4$-branes wrapping the $S^2$ base, pierced by $N$ units of
Ramond-Ramond flux.  When this $S^2$ is large, the low energy on the
unwrapped 2+1--dimensional part of the $D4$-brane preserves two
supercharges, with a field content consisting of a photon, a photino,
a massless neutral scalar and another spinor, where the gauge
half-multiplet is given a topological mass by the presence of a
Chern-Simons term of level $N$ \cite{mcsone,mcsrev}.

In calculating the index $\nu_k$ in this system one can identify (and
factor out) the massless fields with the expected 
translational modes of the wall. The remaining massive gauge modes
can be dualized to massive scalars which decouple, with the caveat
that the Chern-Simons term induces a remnant 
set of quantum mechanical zero modes described by a Landau system, 
\be
 S_{L} = \int dt \left[\frac{1}{2e^2}\dot{A}_i^2 - \frac{N}{8\pi}\ep_{ij}
A_i \dot{A}_j\right], \label{landau}
\ee
Here $A_i=A_i(t)$ ($i=1,2$) are the spatially homogeneous modes of the
gauge field.  The number of vacuum states,\footnote{\,As an aside, this
domain wall system provides an interesting viewpoint on the subtle vacuum
structure of Maxwell-Chern-Simons theory \cite{mcsone,Wnotes}.}
given by the degeneracy of the lowest Landau level
\cite{mcsone,mcsrev}, is $N$ provided the theory has an infrared
regulator on the worldvolume. Consequently, one concludes that the
degeneracy of 1-wall multiplets is $N$ \cite{av}.

This discussion was extended \cite{av} to $k$-walls 
by wrapping $k$ $D4$-branes around the $S^2$. The simplest 
twisted theory arising at low energies is then \none Chern-Simons 
Yang-Mills (CSYM) theory in 2+1D with gauge group U($k$). The massless sector
has now expanded to include a real scalar transforming in the adjoint of
$U(k)$, whose eigenvalues we naturally associate with the positions
of the constituent 1-walls. All but one of these fields should gain
a mass if the configuration is to form a bound state, but the required
mechanism was not apparent in this construction. 
Nonetheless, it was argued that
provided this lifting took place, vacua would arise as in the 
corresponding \ntwo CSYM theory \cite{witten99,csindex} at the 
origin of the moduli space,
with the index taking the value
\beq
\nu_{k} = \left(\begin{array}{c} N \\ k \end{array}\right) 
   \equiv \frac{N!}{k!(N-k)!}\,, 
\label{deg1}
\eeq
which reduces to $N$ for $k=1$ as above. Once again an infrared
regulator on the worldvolume was a necessary condition.

One may view the picture arising from this construction not just as a
quantitative prediction for the spectrum of BPS walls (\ref{deg1}),
but also as an interesting conjecture about the structure of the
worldvolume theory, and it is interesting to contrast it with
expectations from field theory. One unresolved issue
is the existence of additional moduli associated with the constituent 1-wall
positions in a $k$-wall system, as noted above. Given the suggestions that
the effective wall cutoff may scale with $N$ in SYM theory, one possible means of
reconciling these descriptions is to take the large $N$ limit, where
the binding energies are suppressed. This is indeed the expected regime
where the geometric transition is induced \cite{vafadual}.
Having this regime in mind, we will find a worldvolume description within 
SQCD which matches surprisingly well with this construction.

A second tractable deformation of \none SYM theory, which will also
provide a useful reference point, involves compactification
on ${\R}^3\times S^1$ \cite{ahw,sw}. In this process, the 
SU($N$) gauge symmetry can be broken to its maximal Abelian subgroup 
by a Wilson line  $\ph^a = \int_{S^1} {\rm d} x^1 A^a_1$, $a=1,\ldots,N-1$, 
associated with the Cartan components of the gauge field $A_\mu^a$ 
along the compact direction.  If the radius of the $S^1$ is much 
smaller than $\La^{-1}$, and $\ph^a \sim 1$, then we arrive 
at a weakly coupled effective U(1)$^{N-1}$ gauge theory in 2+1D. 
Furthermore, on this Coulomb branch the photons can be dualized to 
periodic scalars $\si^a$ \cite{polyakov}, and the 
system is then described as an \ntwo K\"ahler sigma model with target 
space $T^{2(N-1)}/S_{N-1}$ parametrized by the complex fields
\be
 V^a=\ph^a + \ta \si^a\,.
\ee
This moduli space is lifted by a nonperturbative superpotential
generated by 3D instantons (BPS monopole configurations)
\cite{ahw,sw}. Its general form for gauge group SU(N) has been determined in
various ways \cite{ahw,sw,kv,dhkm,r1} and has the structure of a
complexified affine-Toda potential
\be
 {\cal W} \propto \left[\sum_{a=1}^{N-1}e^{-V^a} + e^{2\pi i\ta}
 e^{\sum_a V^a}\right]. \label{todaW}
\ee
This superpotential therefore leads as expected 
to $N$ chirally asymmetric vacua, and
the corresponding condensates may be continued back to
$\R^{3}\times\R^{1}$  as the
complex structure manifest in (\ref{todaW}) was shown to be
independent of the radius of the circle \cite{sw}. In this system,
the wall configurations allowed by (\ref{todaW}) may be counted
individually as there are no additional moduli. Each wall forms a 
single multiplet, and in the $|k\,n\rangle$ sector one finds 
the same overall multiplicity $\nu_k$ \cite{hiv,av} as in (\ref{deg1}), 
which is consistent given that we again have an explicit infrared regulator.

With these approaches in mind, we will follow an alternate 
strategy, deforming \mbox{\none} SYM theory to weak coupling. To this end
we replace SYM theory by SQCD with $N_f \geq N-1$ massive fundamental flavors, 
a route also followed for the exact calculation of the gluino condensate
\cite{SVone}. When the tree-level mass terms are
large compared to the dynamical scale, we return to pure SYM theory in the
infrared. However, taking the masses as small perturbations, we
pass to a Higgs phase where the gauge fields are heavy and the theory
is well-defined in the infrared \cite{ADS,Seiberg}.
The low energy dynamics of this system is in terms of meson 
quasi-moduli $M$, and in the case of $N_f=N-1$ flavors 
one has the standard 1-instanton induced Affleck-Dine-Seiberg (ADS)
superpotential \cite{ADS} in addition to the mass terms,
\be
 {\cal W} = {\rm Tr}(m M) + \frac{(\La_{N-1})^{2N+1}}{{\rm det} M}\,.
 \label{adsW}
\ee
Note that with a diagonal ansatz for the meson moduli, this
superpotential formally coincides with (\ref{todaW}), up to issues
related to the compactification of the target space. Indeed, the structure 
of this superpotential already allows us to infer
that the wall degeneracy will be nontrivial. As recalled in more
detail below, BPS domain walls describe straight line trajectories
in the ${\cal W}$--plane. Thus, for $N=2$ with a single chiral meson 
field, we see from (\ref{adsW}) that the wall trajectories will 
be given by the roots of a quadratic equation. This suggests a
2-fold wall degeneracy, consistent with (\ref{deg1}), 
which was indeed observed in earlier investigations of this 
system \cite{Stwo}. 

While the similarity with (\ref{todaW}) is apparent, the 
Higgs phase approach can now be seen as a way of bridging both the above
regimes, and thus testing some aspects of the M-theoretic
construction. For gauge group SU($N$), the number of meson moduli 
entering (\ref{adsW}) increases as ${\cal O}(N^2)$, and we will find that
the BPS wall solutions exhibit a moduli space coordinatized by 
Goldstone modes which arise from broken flavor symmetries. It will be
convenient to focus on the case $N_f=N$, where the
full quantum flavor symmetry is manifest, and the corresponding 
description of the Higgs phase is discussed in more detail in the
next section. We find that wall solutions break the flavor symmetries
in such a way that the worldvolume theory, after factoring out the
translational mode, is an \none Grassmannian sigma model in 2+1D. The
vacuum states of this theory therefore count the number of
 BPS wall supermultiplets, Eq.~(\ref{wind}). 

We make significant use of the fact that the
index is independent of variations in the K\"ahler potential.
This and holomorphy of the superpotential are sufficient to ensure
that the index is preserved under the flow back to pure SYM theory. 
We will explicitly show in this way that $\nu_k$ is given precisely 
by (\ref{deg1}) {\it provided} a suitable infrared regulator is in
place in full accord with \cite{av}, e.g. compactifying one spatial 
worldvolume dimension. However, when the external infrared
regulator is removed, the status of the wall multiplicity count 
is less clear. 

The layout of the paper is as follows. In Sec.~\ref{higgsing} we discuss
the structure of the Higgs phase in SQCD when $N_f=N$. With this
background in hand, we turn to the calculation of the BPS wall moduli space in 
Sec.~\ref{modspabps}, while the index calculation, and subtleties 
related to the need for an infrared regulator,
are discussed in Sec.~\ref{sec:four}. Some aspects of the Higgs phase 
system with fewer flavors are discussed in Sec.~\ref{sec:five}, and 
we conclude in Sec.~\ref{sec:six} with some additional comments on the 
worldvolume dynamics.

\section{From SYM theory to SQCD}
\label{higgsing}

A convenient tractable deformation of \none SYM theory is obtained by adding 
$N_f=N$ chiral superfields, $Q_f$ and
$\bar{Q}^g$ ($f,g=1,\ldots, N_f$), transforming respectively in the 
fundamental and anti-fundamental representations of 
the gauge group (the gauge indices are suppressed). This matter 
content is sufficient to completely break the gauge symmetry of the theory in 
any vacuum in which the matter fields have a nonzero vacuum
expectation value. One may then integrate out the gauge
fields obtaining an effective description of the meson moduli matrix $M$,
\be
 M_{f}^g = Q_f \bar{Q}^g,
\ee 
which involves a quantum constraint, first discussed by
Seiberg \cite{Seiberg}. On the non-baryonic Higgs branch, which we 
can restrict our attention to here, this constraint takes the form
\be
 {\rm det}\, M = \left(\La_{N}\right)^{2N},
\ee
which defines a manifold of complex dimension $N^2-1$ in 
${\C}^{N^2}$. Here $\La_N$ is the dynamical scale of SQCD with 
$N_f=N$. 

If we introduce a tree-level mass term
\be
 {\cal W}_{\rm tree} = {\tr}\, (m\,M),
\ee
where $m_f^g$ is the mass matrix in flavor space,
then the effective superpotential can be written as
\be
 {\cal W} = {\tr}\, (m\,M) + \la \left[{\rm det}\, M -
(\La_{N})^{2N}\right], \label{Nform}
\ee
where $\la$ is a Lagrange multiplier. There are
$N$ chirally asymmetric vacua of the theory, which lie at
\be
 \langle M \rangle_k = m^{-1}\, \mu\, \La_{N}^{2}\,\om_N^k
\, , \qquad k=0,\ldots N-1\,, \label{vevs}
\ee
where $\om_N^k=\exp(2\pi i k/N)$ is an $N$-{th} root of unity and
we have defined
\be
 \mu = ({\rm det} \,m)^{1/N}\,.
\ee
For the weak coupling Higgs regime to set in we require that, for
some of the meson fields, $\langle M \rangle \gg \La_{N}^{2}$. From
Eq.~(\ref{vevs}) we see that this is only possible when the mass
matrix  $m$ is hierarchical. For example, we can choose the 
mass matrix in block diagonal form, with the mass of the
$N$-{th} flavor $m_N^N$ much larger than all the others,
\be
 m = \left(\begin{array}{c|ccc}
             m_N^N & 0 & \cdots & 0 \\ \hline
             0 & & & \\
             \vdots & & m' & \\
             0 & & & 
           \end{array}\right), \;\;\;\;\; m_N^N \gg (m')_f^{g},\, \quad
           f,g=1,\ldots, N-1\,. \label{mmatrix}
\ee
We can then integrate out the heaviest flavor, which resolves the 
constraint in (\ref{Nform}) with $M_N^N=\La_N^{2N}/{\rm det}\,M'$
(the fields $M_f^N$ and  $M^g_N$ vanish),
leading to the ADS superpotential \cite{ADS} for the theory
with $N_f=N-1$ flavors, as introduced earlier in (\ref{adsW}),
\be
 {\cal W} = {\tr}\, (m'\,M')+ \frac{(\La_{N-1})^{2N+1}}{{\rm det}M'}\,,
\qquad (\La_{N-1})^{2N+1}=m_N^N\,(\La_{N})^{2N}\,,
\label{adsW2}
\ee
where $m'$ and $M'$ refer to the reduced theory. If all remaining
entries in the mass matrix $m'$ are much less than $\La_{N-1}$ then
the $N$ vacua of the theory lie at weak coupling 
and one can reliably calculate the index $\nu_k$ by
counting solutions of the classical BPS equations. 

Thus, we conclude that to ensure the validity of weak coupling 
analysis the mass matrix should be of the hierarchical form
(\ref{mmatrix}) which diminishes the flavor symmetry.  
Recall that the classical K\"ahler potential ${\cal K}$ of the underlying 
$N$-flavor theory is U($N$)$\times$U($N$) symmetric, although only
SU($N$)$\times$SU($N$)$\times$U(1) is realized canonically in terms
of the meson moduli, where ${\cal K}=\tr \, (\ov{M} M)^{1/2}$.
This symmetry is broken by the superpotential (\ref{Nform}) to 
at most SU($N$), when all masses are equal. The mass matrix
(\ref{mmatrix}) breaks this further to SU($N-1$). Nonetheless, 
the reason we have emphasized the $N$--flavor perspective is that
we can restore the maximal SU($N$) flavor symmetry of
the superpotential, despite the presence of the hierarchical mass
matrix (\ref{mmatrix}), by a holomorphic field rescaling.
In practice it is convenient to introduce the dimensionless fields
$X^g_f$, 
\be 
 X = m\, M\, (\mu\La_N)^{-2}.  \label{Xdef}
\ee 
In terms of $X$ the superpotential (\ref{Nform}) is manifestly SU($N$)
symmetric, 
\be 
{\cal W} = \mu \La_N^2\left[\,{\tr} X + \la \left( {\rm
      det}\, X - 1\right)\right]\,.
\label{Xsup}
\ee
Of course, this is at the expense of diminishing the symmetry of 
the K\"ahler potential, which in terms of $X$ becomes 
${\cal K}\propto \tr\, [ \ov{X} \bar m^{-1} m^{-1} X]^{1/2}$.

The crucial point, as emphasized above, is that the CFIV index
$\nu_k$ does not depend on nonsingular deformations of the 
K\"ahler potential \cite{cfiv}. Thus it is convenient to choose 
a field rescaling that maximizes the symmetry of the superpotential,
as we can then deform the metric back to a more symmetric form
${\cal K}\propto \Tr(\ov{X} X)^{1/2}$ if so desired, for the
purpose of analyzing the BPS equations. Formally, this procedure
is equivalent to taking a symmetric mass matrix in (\ref{Nform}).
However, we emphasize that our dynamical model is that of
(\ref{mmatrix}) and the deformation described above 
is appropriate only for calculating an invariant quantity like $\nu_k$.

\section{Moduli space of BPS Domain Walls}
\label{modspabps}

Having deformed \none SYM theory to SQCD in the Higgs phase, we can 
again verify that the
vacuum structure and supersymmetry algebra still imply the
existence of 1/2--BPS domain walls with tension
determined by the central charge. However, the expression 
for the central charge operator itself is modified \cite{CS}. Ignoring 
total superderivatives, and making use of the Konishi 
relation \cite{konishi}, this operator can be written in the form
form \cite{DSone,Stwo,sv},
\be
 {\cal Z} = \int\! {\rm d} z \,\frac{\partial}{\partial z}\,
 \left\{2\widehat{\cal W }\right\}_{\theta=0}, \qquad
 \widehat{\cal W} = {\cal W}_{\rm tree} 
 -  \frac{T_G- \sum_f T(R_f)}{16\pi^2}\,{\rm Tr}\,W^2\,,
\label{achgt}
\ee
where $z$ is again the transverse coordinate to the wall. The 
first term is due to the tree level superpotential, while 
the second represents the anomalous contribution, given in pure SYM theory
by Eq.~(\ref{symcc}). For $N_f=N$ flavors, which is our main focus here,
the anomalous term vanishes since $T_{{\rm SU}(N)}=N$, and $ T(R_f)=1\,$.

In the Higgs regime at weak coupling this expression reduces in the
$|k\,n\rangle$--sector to the simple form
\be
 {\cal Z}_{k\,n} = 2\left[\,{\cal W}_{k+n} - {\cal W}_{n}\,\right] ,
\label{ZW}
\ee
where ${\cal W}$ is now the effective superpotential (\ref{Xsup}) 
depending on the moduli $X$ while ${\cal W}_{k}$ is the 
value of this superpotential in the $k^{th}$ vacuum,
\begin{equation}
X_k=\omega^k_N \cdot \Id\,,\qquad {\cal W}_{k}=N\mu \La_N^2
\,\omega^k_N\,,
\end{equation}
where $\Id$ is the $N\times N$ unit matrix. For reference, the 
explicit expression for the central charge is given by
\be
 {\cal Z}_{k\,n}=|{\cal Z}_{k\,n}|e^{i\ga_{kn}}=4iN\mu\La_N^2 
\sin\,\frac{\pi k}{N}\exp\left(\frac{i\pi(2n+k)}{N}\right)\,,
\label{ZW1}
\ee
which leads to Eq.~(\ref{tension}) in pure SYM theory, via the decoupling
relation $16\pi^2 \mu\La^2_N \rightarrow 3N \La^3$ as $\mu\rightarrow
\infty$.

BPS walls in this system satisfy  the first order differential 
equations \cite{fmvw,at,cv},
\be
 g_{{\bar a}b}\, \ptl_z X^b = e^{i\ga}\, \ptl_{\bar a}\ov{{\cal W}},
   \label{bpseqn}
\ee
where we have chosen a
convenient basis to expand the meson matrices, in which
the K\"ahler metric is given by 
$g_{\ov{a}b}=\ptl_{\ov{a}}\ptl_{b} {\cal K}\,$,  
and the derivatives $\ptl_{\ov{a}}$ and $\ptl_{b}$ are taken over $\ov{X}$ 
and $X$. Finally, $\gamma\equiv\gamma_{k\,n} $ is the the phase of the
central charge ${\cal  Z}_{k\,n}$ as defined in (\ref{ZW1}).
An important consequence of the BPS equations is that
\be
 \ptl_z {\cal W} = e^{i\ga}\,\|\ptl_i X\|^2\,, \label{Wline}
\ee
and thus the domain wall describes a straight line in the ${\cal W}$-plane
connecting the two vacua \cite{fmvw,at,cv}. 

In calculating the number of solutions to these equations, with
specified boundary conditions in the $|k\,n\rangle$ sector, we will
need a more precise characterization of the dependence of the CFIV
index on variations of the superpotential itself. The structure of the
BPS mass spectrum implies that changes can occur only if a marginal
stability condition is satisfied -- where three vacua align in the
${\cal W}$--plane \cite{fmvw,at}. This allows considerable freedom
(beyond that of deforming the K\"ahler metric) in perturbing the
system in order to verify the existence or otherwise of BPS wall
solutions.

We will make use of this freedom as follows, following the construction
of Cecotti and Vafa \cite{cv}. Firstly, since we consider massive vacua, 
one can expand  the superpotential to quadratic order about each
vacuum, and the set of all linearized solutions to (\ref{bpseqn})
forms a cycle $\De_j$ in field-space diffeomorphic to
a sphere. It then follows \cite{cv} that the weighted soliton number,
namely the index $\nu_k$, is given by the intersection number
of the cycles associated with the two vacua \cite{cv}
\be
 \nu_{k} = \De_n \circ \De_{n+k} \label{int}
\ee 
This formulation of the index is manifestly topological, and provides
a clear picture of how robust it is under deformations. In particular,
it follows from (\ref{int}) that $\nu_k$ counts all trajectories in
the punctured ${\cal W}$-plane (with the vacua excised) which are
homotopic to the straight line connecting the vacua describing the
exact wall solution.  Moreover, the change in the soliton spectrum as
paths wrap around other vacua, crossing curves of marginal stability
via changes in the mass parameters for example, can also be understood
as the intersection numbers change in such a process according to
Picard-Lefschetz monodromies \cite{cv}.

This point of view will prove useful in analyzing the BPS equations
below. However, one must bear in mind that this approach refers
strictly to 1+1D, and thus to a compactification of SQCD on a torus $T^2$.
The stability of the index under decompactification must also
be addressed for any direct application of the results to 3+1D.
Taking these questions in turn, for the remainder of this section
we analyze the space of wall solutions in SQCD, which we will
demonstrate includes continuous flavor moduli, and then move to
a discussion of $\nu_k$ itself in the following section.

\subsection{Broken symmetries and Goldstone modes}
\label{sec:3a}

As we will now demonstrate, the BPS wall solutions in this theory possess
a nontrivial bosonic moduli space ${\cal M}$.
In fact, on the general grounds that a $k$-wall spontaneously 
breaks translational invariance, we have the isometric decomposition,
\be
 {\cal M} = \R \times \widetilde{\cal M},
\label{decomp}
\ee
where the factor $\R$ reflects the center of mass position $z_0$, while
$\widetilde{\cal M}$ denotes the reduced moduli space. The consistency
of this decomposition with supersymmetry can be made explicit if we
lift ${\cal M}$ to the corresponding supermanifold which also encodes
the fermionic moduli. In particular, the two fermionic moduli 
associated with $z_0$ lift $\R$ to the supermanifold $\R^{1;2}$;
quantization of these moduli naturally explains the two state
multiplet structure, described algebraically in Sec.~I, from the
semiclassical point of view. A final point to emphasize is that, since
only two supercharges are realized on the moduli, ${\cal M}$ is a
real manifold, not endowed with any K\"ahler structure.

The decomposition (\ref{decomp}) implies that each $k$-wall
supermultiplet corresponds to a unique vacuum on $\widetilde{\cal M}$.
Therefore, provided we decouple moduli associated with the
translational zero mode, the problem of calculating $\nu_k$ is reduced
to that of finding the Witten index of the worldvolume theory on the
wall. Consequently, we now turn to the problem of deducing the
structure of $\widetilde{\cal M}$.  Given that the low energy
description of SQCD outlined in Sec.~\ref{higgsing} is of
Landau-Ginzburg form, it is natural to expect that this space is
determined purely by the flavor symmetries broken by the wall. In the
remainder of this subsection, we will present the basic symmetry
argument which determines $\widetilde{\cal M}$. In the following
subsection, we show how this arises from a more direct analysis of the
BPS equations.

As reviewed in the previous section, 
the superpotential  (\ref{Xsup}) exhibits a SU($N$) global symmetry which
is preserved by the vacua and is also supported by the deformed 
K\"ahler metric. This superpotential depends 
only on the  eigenvalues $\{\et_i\}$, $i=1,\ldots,N$ of
the dimensionless meson matrix $X$.  In terms of these 
eigenvalues it takes the form
\be
 {\cal W} = \mu \La_N^2 \left[\sum_{i=1}^N \et_i + \la
\left(\prod_{i=1}^N\et_i -1\right)\right].
\label{eigensup}
\ee

Now consider a $k$-wall trajectory, choosing the $|k\,0\rangle$ sector
for simplicity.  The $N$ eigenvalues are restricted to be the same in
each vacuum, i.e. $\et_i=1$ at $z \to -\infty$ and $\et_i=e^{2\pi
ik/N}$ at $z \to +\infty$ for all $i=1,\ldots, N$. 
The modulus of the field is unity in each
vacuum and thus the only `pseudo-topological' means of
characterizing the eigenvalue trajectory in the wall is via its winding
number,
\noindent
\be
  w_i = \frac{1}{2\pi i} \int_{\Ga} \frac{d\et_i}{\et_i},
\ee
where $\Ga$ denotes the wall trajectory.
Up to integer multiples this is clearly $k/N$ for the 
$k$-wall, but we can also keep track of the additional windings. In
principle there are an infinite number of possibilities, but its
clear that energetically only two will occur in stable walls,
namely,
\begin{equation}
w_1=k/N\,,\quad  w_2=(k/N)  -1 
\end{equation}
as exhibited in Fig.~1.
\firstfigfalse
\begin{figure}[h]
 \centerline{%
   \psfig{file=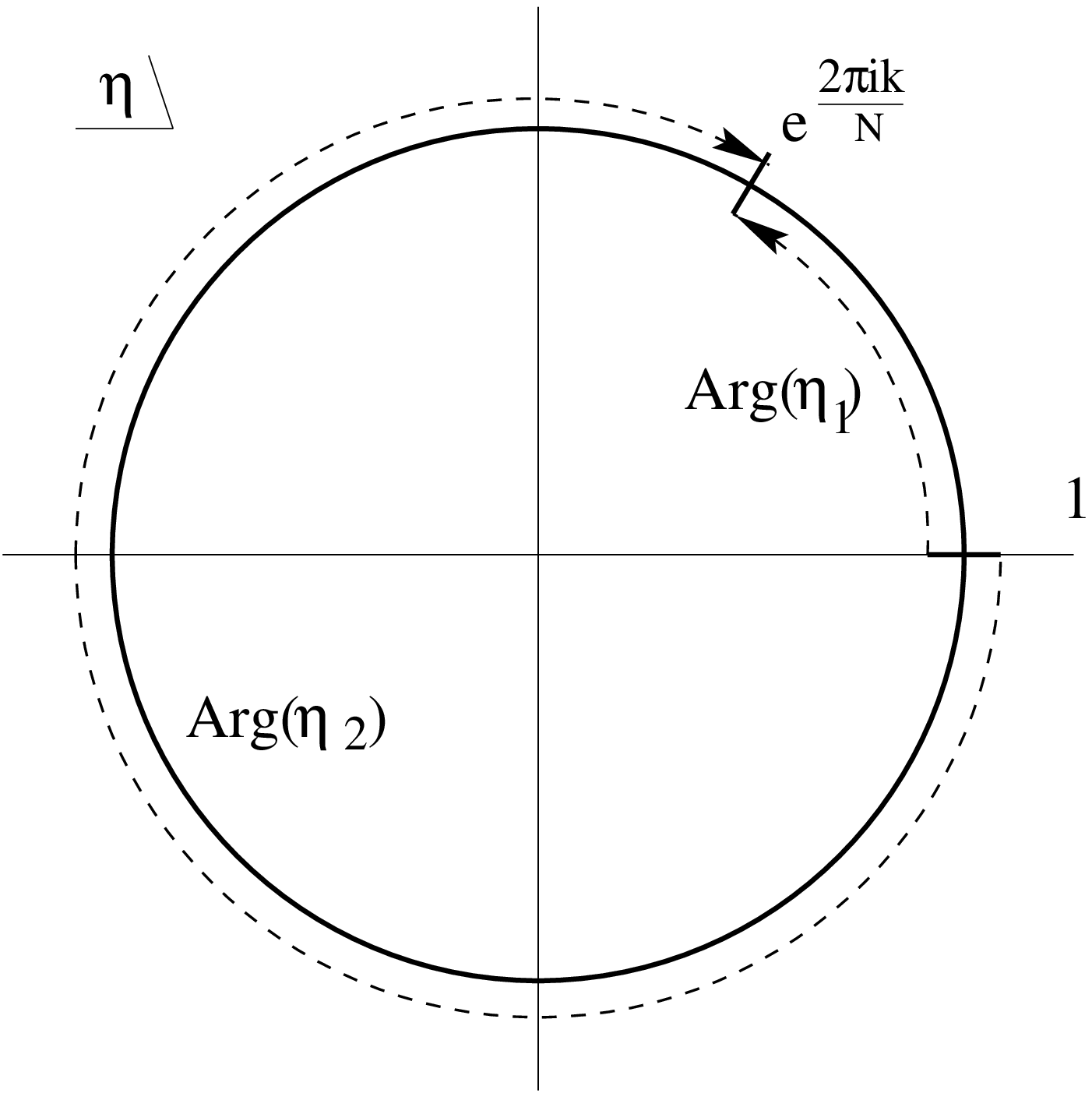,width=6cm,angle=0}%
         }
\newcaption{\footnotesize
 Possible winding trajectories $\et_1$ and $\et_2$ for the eigenvalues.}
\end{figure}
The index is independent of the K\"ahler potential and thus
one can use the freedom to perform diffeomorphisms of the K\"ahler
metric to set all fields characterized by the
same winding numbers equal to each other. To see that this is possible
note that two trajectories $\et_a$ and $\et_b$ in a given
wall with the same winding, say $w_1$, can be mapped into each other
by an analytic mapping on a domain given by the complex plane 
with the vacuum points deleted. In other words from this 
point of view there are only 
two inequivalent trajectories in the wall, up to permutation. 

Now let us prove that for  $k$-wall trajectories there are precisely 
$k$ eigenvalues that have winding 
$w_1$ and $N-k$ having winding $w_2$. Thus, if $N(w_i)$ is the number of
fields having winding $w_i$, then
\begin{equation}
N(w_1)=k\,,\qquad N(w_2)=N-k\,.
\end{equation}
To verify this, denote the trajectories with the 
windings $w_1$ and $w_2$ by  $\et_1$ and $\et_2$, respectively.
If $N(w_1)=l$, the superpotential (\ref{eigensup})
 takes the form
\be
 {\cal W} = \mu \La_N^2\left\{l\et_1 + (N-l)\et_2 +\la\left[\et_1^l
\et_2^{N-l}-1\right]\right\}.
\ee
The constraint imposed by the Lagrange multiplier 
ensures that $\et_1=\et_2^{-(N-l)/l}$. The vacuum values of
the fields then require $k=nl$ with $n\in\Z$, and under 
the assumption that the fields do not undergo multiple windings in the 
wall, as argued above, then we have $k=l$. 

Consequently, since we have two sets of field configurations in the
wall with respectively $k$ and $N-k$ coinciding eigenvalues,
we find that $k$-wall trajectories preserve the following (continuous) 
subgroup of the SU($N$) flavor symmetry:
\be
 {\rm SU}(k) \times {\rm SU}(N-k)\times {\rm U}(1)\,. \label{resid}
\ee
Note that the rank of symmetry group is preserved, following from the
fact that the constraints on $\{\et_i\}$ do 
not break the Cartan torus of SU($N$). Thus far, we have not 
kept track of the global structure of the symmetry groups. 
However, to determine the precise coset associated with the 
broken symmetry generators, and thus the Goldstone modes induced on the
wall worldvolume, it is sufficient to realize that, since our
description involves only adjoint--valued meson fields, the center
of each SU($p$) symmetry group, for $p=N,k$ or $N-k$, acts trivially. 
Thus, the nonabelian symmetry groups are strictly of the form 
SU($p$)$/\Z_p$. 

Therefore, taking this global structure into account, we deduce that
the reduced moduli space for $k$-walls is the complex 
Grassmannian\footnote{\,We thank A.~Smilga for helpful discussions on 
the geometry of soliton moduli.} 
\be
 \widetilde{\cal M}_k = G(k,N) \equiv 
  \frac{{\rm U}(N)}{{\rm U}(k) \times {\rm U}(N-k)}
\ee
reflecting the Goldstone modes induced by the broken flavor
symmetries. For 1-walls, $G(1,N) = \C{\rm P}^{N-1}$.
This result apparently depends on the number of flavors, but
can nonetheless be used to determine the worldvolume Witten index
as we shall discuss in more detail below.

One may note that the information on eigenvalue trajectories is
actually sufficient to deduce the `classical' wall degeneracy.  It is
given by the number of possible permutations of the eigenvalues (as
discussed in \cite{hiv}), leading to the result for $\nu_k$ given in
(\ref{deg1}). However, this argument neglects infrared effects on the
wall, an understanding of which requires a more detailed analysis of
the worldvolume dynamics. Before turning to this, we will present in
the next subsection a more explicit derivation of $\widetilde{\cal M}$
following from the BPS equations.

\subsection{Analysis of the BPS equation}

The BPS equations (\ref{bpseqn}) depend not just on the form of the
superpotential, but also on the K\"ahler metric, and it is the latter
dependence which, while irrelevant for obtaining $\nu_k$, 
determines the precise wall profile. Therefore, it is
useful to disentangle the crucial dependence on the moduli, arising
from the superpotential, from the inessential details relating to
the actual profile of the wall.

To this end, we first introduce a basis of Hermitian matrices
$\{T_0,T_A\}$, $A=1,..., N^2-1$, where $T_0=\Id/\sqrt{N}$ and 
$\{T_A\}$ are orthonormal generators for the Lie algebra of  SU($N$), 
satisfying $\tr\,(T_A T_B) = \de_{AB}$.
Then we can expand the dimensionless meson matrix in this basis,
\be
 X = \sqrt{N}\left(X^0 T_0 + iX^A T_A\right), \label{basis}
\ee
where $\{X^0,X^A\}\in\C$. In this basis, the vacua lie at 
\be
 \langle\, X^0\,\rangle = \om_N^k\,, \qquad
 \langle \,X^A \,\rangle = 0\,, \label{vac}
\ee
and on imposing the constraint obtained by integrating out the Lagrange
multiplier $\la$, the superpotential takes the form
\be
 {\cal W} =\mu \La_N^2\, X^0\,. \label{W0t}
\ee
BPS wall profiles are then given formally by parametrizing the constraint that
${\cal W}$, and thus $X^0$, 
must follow a straight line connecting the two vacua, as follows from
Eq.~(\ref{Wline}).  In other words,
{\it any} $k$-wall trajectory, parametrized by
a fiducial scale $t\in[0,1]$, is given by the relation
\be
 X^0(t) = f(t) \equiv (1-t) + t\om_N^k\,, \label{X0t}
\ee
with $X^A$ subject to the constraint
\be
 {\rm det}\left[X^0(t)\Id + i\sqrt{N}\,X^AT_A\right]=1
 \label{ms}
\ee
with the appropriate boundary conditions at the vacua. The existence, 
and the moduli space, of wall solutions
then devolves on the analysis of this constraint.

In essence, we have simply shifted the nontrivial field dependence
from the superpotential to the metric. With a suitable coordinate
choice the BPS equations will then
provide a nontrivial profile for a single coordinate, consistent with
Eq.~(\ref{Wline}), with the other coordinates either fixed or
remaining constant and thus leading to moduli. We turn first to the
simplest case with gauge group SU(2).

\subsubsection{Gauge group SU(2)}

We will study first the theory with gauge group SU(2), and two
flavors. In this case, the constraint (\ref{ms}) takes
the form
\be
 \sum_{a=0}^3 (X^a)^2 = 1\,, \label{cone}
\ee
where $a=(0,A)=0,\ldots, 3$. This constraint yields
a smooth complex submanifold of $\C^4$, known as the 
deformed conifold.\footnote{\,The singular conifold is
recovered in the classical limit, $\La\rightarrow 0$, where
Eq.~(\ref{cone}) reduces to $\sum_{a=0}^3 (X_{\rm class}^a)^2 = 0$
with $X_{\rm class}=\lim_{\La\to 0} (\La^2 X)$.}

The real section of the deformed conifold (\ref{cone}) defines a
3-sphere, and we see that the two vacua of the theory given by
(\ref{vac}) lie at antipodal points on this $S^3$. The geometry
of the space becomes more transparent by defining a radial coordinate $r$,
\be
 r^2 = {\tr}\,(\ov{X}X)\,,
\ee
where constant values of $r$ define a foliation with sections 
having the generic form SU(2)$\times$SU(2)/U(1), where the U(1) acts
in such a way that this space is 
topologically $S^2 \times S^3$, see \cite{Candelas}. 
However, at the minimal radius $r=1$ 
the coset becomes SU(2)$\times$SU(2)/SU(2)
and these sections collapse to an $S^3$ identifiable as the real
section above. Thus, the manifold is conical for large $r$, while 
the apex of the cone is rounded off to an $S^3$.

Supersymmetry demands that we use a K\"ahler metric on the deformed
conifold. Such a metric preserving the SU(2)$\times$SU(2) action
apparent from the $r\neq 1$ sections will be a function only of $r^2$.
In terms of the K\"ahler potential $K=K(r^2)$, and the meson matrix $X$, the
metric will take the form
\be
 ds^2_C = K'(r^2) \,{\tr} \,(d\ov{X} dX) + 
 K''(r^2)\,|{\tr}\, (\ov{X} dX)|^2\,. \label{metdc}
\ee
K\"ahler metrics on the deformed conifold, which are also Ricci flat,
were first obtained in \cite{Candelas}. Supersymmetry does
not impose the latter constraint here,\footnote{\,In contrast, with \ntwo SUSY 
the Higgs branch metric is required to be hyperK\"ahler, which would imply
Ricci-flatness.} but in fact we will not need to
choose a precise form for $K(r^2)$ away from the apex. In this region, the
symmetry we have imposed is sufficient to ensure that the metric 
(\ref{metdc})  takes the asymptotic form \cite{Candelas,defcon}
\be
 ds^2_{r\rightarrow 1} =\La_2^2 \left[ d\rh^2 + d\Om_3^2
         + \frac{1}{2} \rh^2 d\Om_2^2\right],
\ee
where $\rh=\sqrt{2(r-1)}$ and 
$d\Om_2^2$ represents a 2-surface (topologically $S^2$) which shrinks
as $\rh\rightarrow 0$, while $d\Om_3^2$ is a round $S^3$ -- equivalent
to the real section -- which remains with finite volume at the apex. 
Thus, the metric reduces locally near $r=1$ to ${\R}^3\times S^3$.

The crucial simplifying feature is that it is not just the vacua, but
the entire wall trajectory which lies on the real section. To see this
we adopt a different viewpoint on the deformed
conifold, first discussed by Stenzel \cite{stenzel,stenmet}. 
The manifold is symplectic, given by the co-tangent bundle $T^*(S^3)$
over $S^3$,  and this is made manifest by introducing 
real coordinates $x^a$ and momenta $p_a$ via
\be 
 X^a = \cosh(\sqrt{p_bp_b}) x^a +
 i \,\frac{\sinh(\sqrt{p_bp_b})}{\sqrt{p_c p_c}}\, p_a\,, 
\qquad a,b,c =0,1,2,3\,.
\ee
The defining constraint then takes the form,
\be
 \sum_{a=0}^3 (x^a)^2 = 1\,,\qquad \sum_{a=0}^3 x^a \,p_a =0\,,
     \label{symp}
\ee
which describes the canonical phase space of a dynamical system
with configuration space $S^3$, with momenta lying in the co-tangent
space. 

Now, from the condition that the superpotential lie along a real line
connecting the two vacua, see Eqs.~(\ref{W0t}), (\ref{X0t}), 
we deduce that $p_0(t)=0$, and the superpotential reduces
to ${\cal W} = \mu\La_N^2 \cosh(\sqrt{p_Ap_A})x^0$, where $A=1,2,3$. Since
$\cosh(\sqrt{p_Ap_A})$ is strictly positive, and the 
vacua lie on the real section so that $p_A(0)=p_A(1)=0$, 
we can smoothly deform the K\"ahler potential if 
necessary so that $p_A(t)=0$, $A=1,2,3$, for all $t$. 
One can readily verify that this solves the constraints (\ref{symp}). 
The solution then corresponds to a zero ``energy'' configuration 
of the analog system, where $E=\sum_{a=0}^3 (p_a)^2/2$, and 
remains entirely within the real section $S^3$ at $r=1$.

By deformation, we have chosen the K\"ahler metric on the 
field theory moduli space to respect the maximal SU(2)$\times$SU(2)
symmetry, and so the corresponding metric on $S^3$
will be the round one. Re-expressing the real coordinates 
$\{x^a\}$, $a=0\ldots 3$, on $S^3$ in terms of Euler angles
$\{\th,\xi,\ph\}$, the metric takes the form
\be
  ds^2_{r=1} =\La_2^2 \,d\Om_3^2\,, \qquad
 d\Om^2_3 = d\th^2 + \sin^2\th(d\xi^2+\sin^2\xi d\ph^2)\,, \label{2met}
\ee
while the superpotential, restricted to the real section, is given by
\be
 \left.{\cal W}\right|_{\rm trajectory} 
 = 2\mu \La_2^2 \cos\th.
\ee 
The vacua lie at the poles $\th=0,\pi$, and the BPS equations
reduce to 
\be
 \ptl_z \th = 2\mu \sin\th\,, \qquad \ptl_z \xi = \ptl_z \ph =0\,,
\ee
which are solved by the sine-Gordon soliton,
\be
 \th(z) = 2\arctan e^{2 \mu (z-z_0) }\,.
\ee
This solution,
schematically represented in Fig.~2, is characterized by 
three moduli $\{z_0,\xi_0,\ph_0\}$. The angular
modes are Euler angles on the 2-sphere, and so we recover the
moduli space deduced earlier from symmetry considerations,
\be
 {\cal M}^{\rm N=2} = {\R} \times {\C}{\rm P}^1,
\ee
where since $N=2$, we have necessarily been considering minimal 
walls with $k=1$.

\begin{figure}[h]
 \centerline{%
   \psfig{file=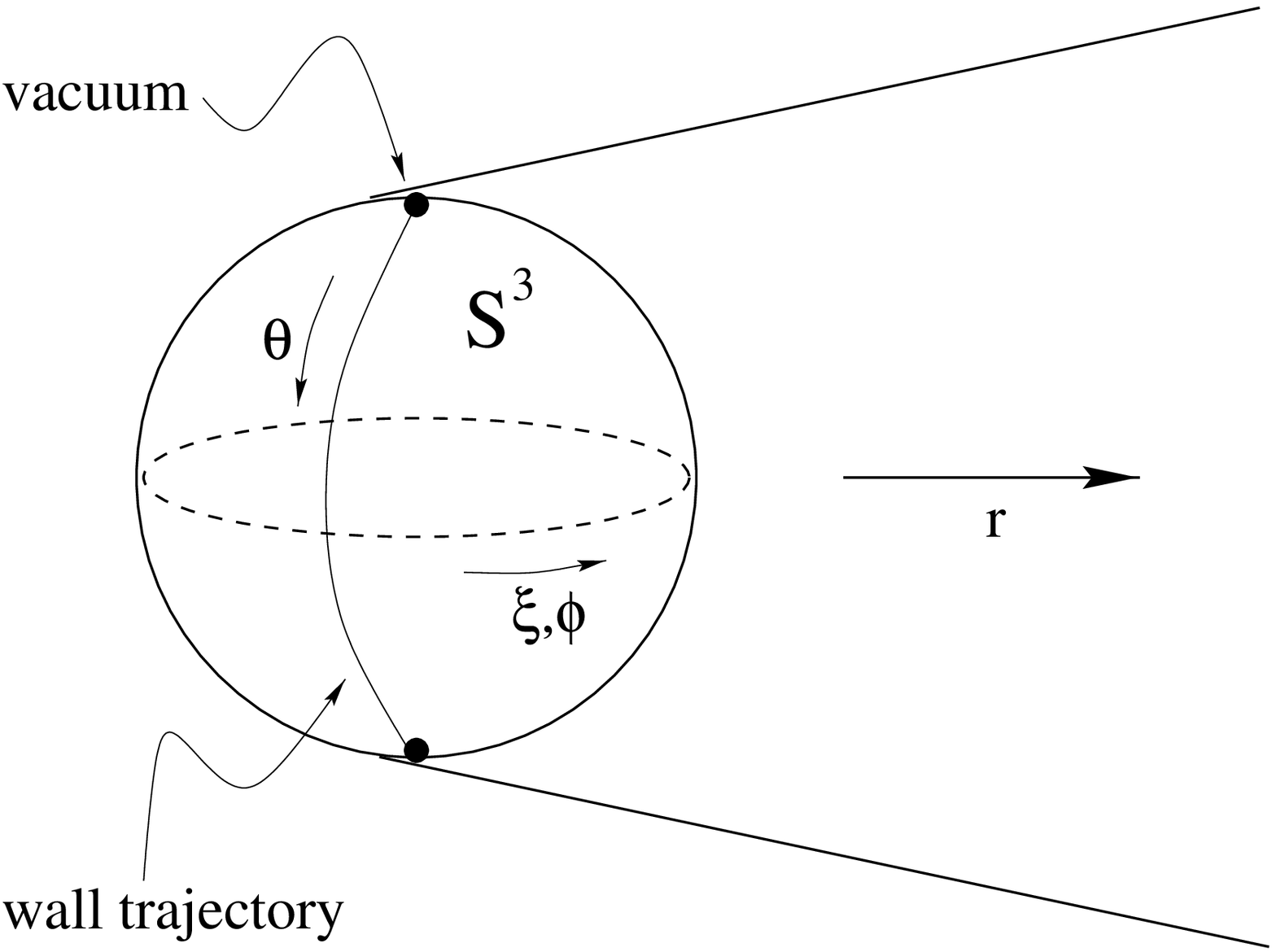,width=7cm,angle=0}%
         }

\vspace*{0.1in}

\newcaption{\footnotesize
 Wall trajectory on the resolved $S^3$ at the tip of the
 deformed conifold.}
\end{figure}
\noindent

In preparation for later analysis, we can compactify the spatial
dimensions on a 2-torus of area $L^2$. Then, at excitation energies
$E \ll L^{-1}$, we are effectively reducing the 
system to one of kinks in 1+1D, where this reduction to quantum
mechanical moduli makes sense. Integrating over the wall
profile leads to the corresponding bosonic moduli space Lagrangian 
which takes the form
\be 
 L_{\rm bose} = -M + \frac{1}{2}\,M \dot{z_0}^2 
+ R_{\widetilde{\cal M}} \left(\dot{\xi}^2 + \sin^2\xi \dot{\ph}^2\right),
\ee
where $M=T_1 L^2$ is the kink mass, determined by the wall tension 
$T_1=8\mu\La_2^2$, while integrating over the wall profile
leads to a scale 
\be
 R_{\widetilde{\cal M}} \sim L^2 \frac{\La_2^2}{\mu} 
    \label{Mscale}
\ee
for the flavor moduli space $\widetilde{\cal M}$.

\subsubsection{Gauge group SU(N)}

The general case can be understood by following a slightly less direct
approach. It will be convenient to describe generic $k$-walls 
as embedded sine-Gordon solitons, although we emphasize that 
this simply reflects a particular choice of the K\"ahler metric. 
In more detail, along the wall trajectory we make the change of variables
$t=\sin^2\th/2$ where $\th\in[0,\pi]$. The
superpotential (\ref{W0t},\ref{X0t}) can then be written as
\be
 \left.{\cal W}\right|_{\rm trajectory} 
 = N \mu  \La_N^2 e^{i\pi k/N}\left[\cos\frac{k\pi}{N}
 -i\sin\frac{k\pi}{N}\cos\th\right]. \label{Wtraj}
\ee
If we now choose a K\"ahler potential of the form
\be
 {\cal K} = \La_N^2 
 \left[\th^2 + {\tr}\,f(\ov{\tilde{X}}\tilde{X})\right], \label{met1}
\ee
where $f$ is any smooth function along the wall trajectory and
$\tilde{X}=X-(1/N)({\tr}\,X)\Id$ 
is the trace-free part of $X$, then the BPS equation
for $k$-walls takes the form
\be
 \ptl_z \th = N \mu  \sin\frac{k\pi}{N}\sin\th,
\ee
which is again solved by the sine-Gordon soliton,
\be
 \th(z) = 2 \arctan e^{2 \tilde\mu (z-z_0) }\,, \qquad 
\tilde\mu =\frac{N}{2}\, \mu 
\sin\frac{k\pi}{N}\,.
\ee

This profile is dependent entirely on the choice of the metric
(\ref{met1}), which naturally generalizes (\ref{2met}). 
However, since we are not concerned with its precise
form the only point of concern is whether it introduces additional
singularities. To check this, we need to compare it with the canonical
metric at weak coupling, which is the only regime in which the metric
is known. One can verify that the mapping to (\ref{met1})
is nonsingular in the vicinity of the wall trajectory. 

With this profile in hand we can now investigate the moduli of this
solution. The corresponding reduced moduli space
$\widetilde{\cal M}$ is determined by the space of smooth solutions to the
constraint (\ref{ms}) with fixed boundary conditions at the appropriate
vacua. 

Geometrically, the manifold (\ref{ms}) is a determinental variety of
complex dimension $N^2-1$, which is rather difficult to analyze
directly, except in the SU(2) case discussed above, and so we will
adopt a different approach motivated by the discussion in
Sec.~\ref{sec:3a}
First of all, note that although the flavor symmetry of the 
theory is SU($N$), the symmetry of the
constraint (\ref{ms}) is its complexification SL($N,{\C}$). 
This allows us to diagonalize the meson matrix $X=\sqrt{N}(X^0T_0+iX^AT_A)$ 
by the adjoint action of SL($N,{\C}$), with $X \rightarrow$
diag$\{\et_i\}$ as before, where now the eigenvalues are 
functions of a single complex variable $x$. Following the arguments
of Sec.~\ref{sec:3a}, to solve the constraint 
\be
 \prod_{i=1}^N \et_i(x(t)) = 1, \label{ms2}
\ee
we must impose $\et_i=\et_j$ for $i,j=1,\ldots,k$, and 
$\et_k=\et_l$ for $k,l=k+1,\ldots,N$. We can
then represent the diagonalization of $X$ in the form
\be
 X \rightarrow f(t)\Id + x\sqrt{N}\,\Om \label{ansatz}\,,
\ee
where we have used (\ref{X0t}) and $\Om$ is the following generator
of the Cartan subalgebra
\be
 \Om = {\rm diag}\left\{-\sqrt{\frac{N-k}{Nk}} \,\Id_k\,,
\;\sqrt{\frac{k}{N(N-k)}} \,\Id_{N-k}\right\},\label{Tdef}
\ee
with $\Id_k$ the $k\times k$ unit matrix.

The constraint (\ref{ms2}) then reduces to 
\be
 \left( f(t) + \sqrt{\frac{N-k}{k}}\,x\right)^k
\left( f(t) -\sqrt{\frac{k}{N-k}}\,x\right)^{N-k} = 1\,.
\ee
The number of solutions $x(t)$ to this equation asymptoting to the 
vacua $x(0)=x(1)=0$ can be obtained as follows \cite{hiv}. Note
firstly that the constraint is resolved by defining a new variable $y$,
\be
 y^k = f(t) - \sqrt{\frac{k}{N-k}}\,x, \qquad
 y^{-(N-k)} = f(t) + \sqrt{\frac{N-k}{k}}\,x\,.
\ee
An apparent phase ambiguity which is dropped in this
transition is actually fake once the vacua are fixed.
Then, eliminating the $x$ dependence we obtain,
\be
 \frac{{\cal W}_{\rm ansatz}}{\mu\La_N^2} = N f(t) = k y^{k-N} + (N-k)y^k, 
\ee
which we recognize as the constraint that the
ADS superpotential, evaluated within the ansatz (\ref{ansatz}),
follows the straight line wall trajectory. Note that another 
phase ambiguity has been dropped to ensure the correct 
asymptotic vacua.  

Since the vacua are massive, and thus the second derivative of the 
superpotential is finite, there are at most two possible 
trajectories emanating from each vacuum point. However, a 
perturbative analysis shows that only one of these can 
interpolate between both. Existence of this unique solution can 
be demonstrated \cite{hiv} by taking the trial solution
$y(t)=e^{2\pi i t/N}$ and showing that its image in the punctured
${\cal W}_{\rm ansatz}$--plane (with the vacua excised) is 
homotopic to a straight line, the latter describing the exact 
wall trajectory.

Thus, we have found precisely one solution for all $k$ associated with
the ansatz (\ref{ansatz}) and the generator $\Om$. Consequently,
following standard arguments, the moduli space of these solutions is
given by the coadjoint orbit of $\Om$ under the symmetry group, which
in this case is SU($N$). This is the manifold swept out by the adjoint
action of SU($N$) mod the stability group of $\Om$, which we see
immediately is SU($k$)$\times$SU($N-k$)$\times$U(1) up to discrete
factors.  Thus, taking into account the fact that the center of each
nonabelian group acts trivially, we recover the result obtained
earlier that the reduced moduli space,
\be
 \widetilde{\cal M}_k = G(k,N)\,,
\ee 
is the complex Grassmannian of $k$-planes in ${\C}^N$.

Having factored out the transverse position modulus, which is
decoupled and not visible in the construction above, we find that the
moduli associated with the broken flavor symmetries induce a
nontrivial 2+1D sigma model on the wall worldvolume, the
supersymmetric vacuum states of which -- to be identified with
inequivalent $k$-walls -- we will count in the next section.

\section{The worldvolume Witten index}
\label{sec:four}

Having determined the structure of the bosonic moduli space as 
$\R \times \widetilde{\cal M}_k$, the calculation of $\nu_{k}$
simplifies in that the massless field associated with the
translational zero mode is factorized, and its associated multiplet
decoupled. We then identify each inequivalent $k$-wall with a unique
vacuum state of the \none sigma model on $\widetilde{\cal M}_k$,
and count then with the Witten index. The index is conveniently 
defined by imposing an infrared regulator on the spatial coordinates 
of the worldvolume. In general, the result is insensitive to removal 
of the regulator if the theory has a mass gap, an issue which requires
some dynamical knowledge of the system in question. In this section,
we discuss the status of the index for the worldvolume sigma model
on $\widetilde{\cal M}_k$ by systematically removing the infrared
regulators. 

\subsection{Compactification on \boldmath{$T^2$}}

We first consider a fully regularized system putting 
the spatial part of the worldvolume on a torus. 
Then in the low-energy limit, the worldvolume theory reduces to 
a quantum mechanical problem with the moduli dependent only on time. 
In effect, we are now analyzing the quantum mechanical moduli of
kinks in 1+1D, as discussed in detail in \cite{cv}. Given that
the flavor moduli parametrize the Grassmannian $G(k,N)$, which is
compact, the techniques for calculating the index and 
thus the number of quantum ground states were described in
\cite{Witten1}, and the result is given by \cite{cv}
\be
 \nu_k = \ch(G(k,N)) = \left(\begin{array}{c} N \\ k \end{array}\right) 
   \equiv \frac{N!}{k!(N-k)!}\,, \label{deg2}
\ee
where $\ch(G(k,N))$ is the Euler characteristic of the Grassmannian.
The resulting spectrum of $k$-walls is consistent with the
results of \cite{av}. Note that the result is independent of the
original SQCD K\"ahler metric, and depends only on the topology of the
bosonic moduli space. This is a necessary consistency check as we have
relied on the independence of the result under smooth deformations of the
metric \cite{cfiv}. Moreover, the invariance of the index 
under small perturbations of the superpotential, for which no 
vacua become aligned \cite{cv}, is now transparent. Specifically,
were we to perturb the meson mass terms slightly, thus reducing the 
residual symmetry of the wall, the spectrum would not change as 
this would deform the metric on $G(k,N)$ but clearly not its topology. 

Let us note that in the context of kinks, the degeneracy (\ref{deg2})
has an interesting interpretation. The moduli space 
for 1-walls is ${\C}{\rm P}^{N-1}$, and the corresponding degeneracy 
from  (\ref{deg2}) is $N$. It is natural to interpret this 
in terms of the walls forming an $N$-plet of SU($N$), which
is the isometry group of ${\C}{\rm P}^{N-1}$ (see also \cite{hiv}). 
The degeneracy (\ref{deg2}) then implies that composite $k$-walls
fall into antisymmetric tensor multiplets of SU($N$), namely the
$k$-th fundamental representation. This implies that 1-walls, when
reduced to kinks in 1+1D are `fermionic' in flavor, consistent with 
expectations for solitons in similar Landau-Ginzburg systems.

\subsection{Compactification on  \boldmath{$S^1$}}

We now decompactify one cycle of the torus.
The index obtained above will remain valid provided vacuum states 
cannot disappear to infinity in the process of decompactification.
In this case, we are left with the moduli dynamics described
by an \none Grassmannian sigma model in 1+1D. Such systems are 
well understood, the nontrivial infrared behavior of the
1+1D sigma model restores the original SU($N$) symmetry,
allowing a dynamically generated mass gap
for the flavor moduli. Since the vacua are massive, we are guaranteed 
that on considering scales below the gap the problem reduces to
one of quantum mechanics as above. Thus we can conclude
that the number of discrete supersymmetric vacua, and therefore the
spectrum of $k$-walls in SYM theory compactified on ${\R}^3\times S^1$
is still given by (\ref{deg2}). This conclusion is clearly consistent with
the results obtained by direct compactification of SYM theory \cite{hiv,av}, 
as reviewed in Sec.~\ref{intro}.

In concluding this subsection, we note that the 1+1D sigma model 
provides another interesting point of view on the degeneracy (\ref{deg2}). 
In particular, the $N$-plet wall multiplet structure seems in this
case rather closely tied to restoration of the SU($N$) 
symmetry in the infrared.

\subsection{Decompactification and an alternative regulator}

On decompactifying the second cycle of the torus, we have removed all
infrared regulators and the status of the index (\ref{deg2}), in as
far as it correctly describes the wall multiplicity, devolves
on the infrared dynamics of the \none Grassmannian sigma model in 2+1D. 
This system, specifically the ${\C}{\rm P}^{N-1}$ model, has
received  less attention than the corresponding models in 1+1D.  
In perturbation theory there is no evidence for infrared divergences,
and this conclusion extends to leading order in the large $N$
expansion \cite{adcf,isy}. However, due to the fact that the flavor modes
can be combined into complex chiral multiplets, there is no obvious 
symmetry \cite{wit95} or anomaly constraint \cite{ahw}
which forbids a mass term. Moreover, one must also bear in mind that the 
UV divergences of the model are cut off physically at a scale 
given by the inverse width of the wall, which is ${\cal O}(m)$ in
SQCD, and this introduces an additional scale.
Thus, without a more detailed understanding of the dynamics of this UV
regularized \none $G(k,N)$ sigma model, the status of the wall
multiplicity count remains unclear after decompactification. 

In contrast, the index itself can be regulated in an alternative manner
via a perturbation of the theory which lifts the additional
flavor moduli.\footnote{\,We thank A.~Losev for suggesting this approach
and for related discussions.}  In practice, we require a perturbation
which lifts the off-diagonal elements of the meson matrix, so that
the system reduces to a theory of the eigenvalues 
$M \rightarrow $ diag$\{\et_i\}$ with 
\be 
 {\cal W} = \mu \La_N^2 \left[\sum_{i=1}^N \et_i + \la
\left(\prod_{i=1}^N\et_i -1\right)\right]. \label{esup2}
\ee
as in (\ref{eigensup}). One may then construct wall solutions which
possess no flavor moduli, and determine the multiplicity directly as in 
\cite{hiv,av}, finding the result (\ref{deg1}) once again.

There are several possible mechanisms for lifting the off-diagonal 
modes and all have certain side-effects. One can simply perturb the
superpotential ${\cal W} \rightarrow {\cal W}+\delta{\cal W}$ with nonlinear
terms $\delta{\cal W}$ which break the SU($N$) flavor symmetry, but this
generically introduces new vacua, and care is needed in 
discarding any spurious wall solutions  
which decouple as the perturbation is removed. 
Alternatively, one may
weakly gauge the flavor symmetry, under which the meson matrix $M$
transforms in the adjoint. The decoupling limit for the gauge modes
then enforces a $D$-term constraint, $\Tr\,[M,\overline{M}\,]^2 = 0$, 
ensuring that
the off-diagonal modes of $M$ are lifted, leading to 
(\ref{esup2}). This procedure does introduce an additional decoupled 
set of light U(1) fields, but has the merit of
retaining the symmetry structure we expect to be important in the pure
SYM regime, as is apparent on comparison with the approach of
compactifying on ${\R}^3\times S^1$, see Eq.\,(\ref{todaW}).

Perturbing the original theory in this way indicates that
the index is, as it should be, stable under different choices 
for the regulator. However, its connection to the physical wall 
multiplicity still rests on the question of stability under
removal of the regulator. As we noted above, this can be rephrased 
as a dynamical question about the vacuum structure of the worldvolume
sigma model. On this issue, we will limit ourselves here to a few comments 
describing the two possible scenarios. 

The first is that a nonperturbative mechanism generates a mass
for the flavor modes, which implies that the index remains unchanged.
In this regard, recall that supersymmetric nonlinear sigma models are most 
conveniently studied by embedding them in a corresponding
gauged linear sigma model \cite{wit93,wit94}. For the
${\C}{\rm P}^{N-1}$ model, this gauge theory is Abelian and 
at first sight there are no obvious non-perturbative effects which could 
generate a mass gap. However, the presence of the UV cutoff
complicates this issue, as the UV completion of the theory may allow
nonperturbative mass generation. A classic example, although not
directly relevant here, is the Polyakov mass for the photon 
\cite{polyakov,ahw} in U(1) theories where, from the low energy 
perspective the nonperturbative mechanism  involves `singular'
instantons, which are resolved above the cutoff scale. 

With this in mind it is intriguing to note that, if we assume for 
a moment that a mass gap for the flavor modes were to arise via some
mechanism, one could integrate them out within
the linear sigma model, which in effect corresponds to flowing back to pure
SYM theory. This process is known to induce a standard 
kinetic term for the gauge fields, and in 2+1D will also lead to 
a Chern-Simons term \cite{adcf,isy}. In the case of 1-walls,
the resulting system would be \none Maxwell-Chern-Simons at level $N$
(up to higher derivative corrections), remarkably consistent 
with the worldvolume theory obtained by Acharya
and Vafa \cite{av}. This connection can also be made, at a formal
level, within the compactified system, where the light 
flavor fields have masses scaling inversely with the volume. 

The second scenario is that the flavor modes remain massless at the
quantum level, and a priori there is no obvious inconsistency with 
this. More precisely, as we flow back to \none SYM theory,
these modes must disappear, but this can occur without direct mass
generation. In particular, the flavor modes may `freeze'
in this limit. Recall for the SU(2) case, that the K\"ahler class of
the 2+1D $\C$P$^1$ sigma model scales as
\be
 M_{\widetilde{\cal M}} \propto \frac{\La_2^2}{\mu} = 
 \frac{3\La^3}{8\pi^2\mu^2}\,,
\label{Kclass}
\ee 
where $\La$ is the \none SYM scale defined in (\ref{gcond}) 
and $\mu=({\rm det}\,m)^{1/N}\,$.
Thus, as we begin to decouple the matter fields, the moduli space 
shrinks. In the absence of any mass generation, 
it is plausible that in the limit that $\mu \rightarrow \infty$, 
the manifold $\widetilde{\cal M}$ shrinks to zero size, and 
the corresponding flavor modes are frozen, and consequently decouple. 
However, this conclusion requires a significant assumption 
about the behavior of the SQCD K\"ahler metric in this regime. 
It would be natural for the scale (\ref{Kclass}) to have corrections 
of ${\cal O}(\mu/\La)$ which may significantly change its form 
when $\mu \geq \La$.

\section{Breaking the flavor symmetry and \boldmath{$N_f=N-1$}}
\label{sec:five}

Thus far, we have purposefully chosen to work with $N$ flavors,
to make the multiplet structure of the walls under global
symmetries quite explicit. This required us to make use of the
invariance of the index under $D$-term deformations, so as to restore
the maximal SU$(N)$ symmetry of the $N$-flavor theory.
Recall that in Sec.~\ref{higgsing}, for consistency we 
required that the mass matrix was hierarchical to ensure a 
weak coupling regime, which implies that the weak coupling flavor
symmetry of the underlying theory is at most SU$(N-1)$.

While this approach was convenient for obtaining the index, it is also
instructive to see how the explicit breaking of flavor symmetries is
manifest at the level of the flavor moduli space $\widetilde{M}$, in a
regime where the direct relation to Goldstone modes is lost.  For ease
of illustration, we focus on the simplest case with gauge group SU(2).
In the analysis of Sec.~III, as just described, we deformed the
K\"ahler metric so as to restore the maximal SU(2) symmetry of the
theory, despite the hierarchical mass matrix of Eq.~(\ref{mmatrix}),
\be
 m = {\rm diag}\{m_1,m_2\},\qquad m_1 \ll m_2 \ll \La_2.
\ee
At the level of the K\"ahler potential, setting $m_1 \neq m_2$ breaks
the symmetry from 
SU(2)$\times$SU(2)$\times$U(1)$\,\cong\,$SO(4)$\times$SO(2) to 
SO(2)$\times$SO(2).

This reduction in symmetry can be traced through to the metric
structure of the wall moduli space as follows. Recall that in the
analysis of Sec.~III, the wall profile described a path between 
the poles of an $S^3$ -- the real section of the deformed conifold -- 
with the $S^3$ having the round SO(4)--invariant metric. 
It is convenient to visualize the $S^3$ via stereograhic projection 
as $\R^3 \cup \{\infty\}$. The poles of the $S^3$ and thus the 
two vacua of the theory are projected to 0 and $\infty$, and hence 
the wall trajectory is described by a 
line from the origin to infinity in $\R^3$. This line is  
parametrized by Euler angles on the sphere at fixed radius, and hence this
construction realizes the $\C$P$^{1}\cong S^2$ flavor moduli space as a
submanifold of $\R^3$. In Sec.~III, the SO(4) symmetry of the 
original K\"ahler metric ensures an SO(3)--invariant round
metric on the $S^2$. However, with unequal masses, the metric induced
on the $S^2$ has only an SO(2) isometry, which we can arrange to
generate rotations in the horizontal plane. Metrically, the $S^2$
moduli space is then an ellipsoid, with the ellipticity characterized
by the dimensionless parameter,
\be
 \xi \,\equiv\, \frac{m_2-m_1}{\sqrt{m_1m_2}}.
\ee
This metric structure for the moduli space is exhibited in Fig.~3.

\begin{figure}[h]
 \centerline{%
   \psfig{file=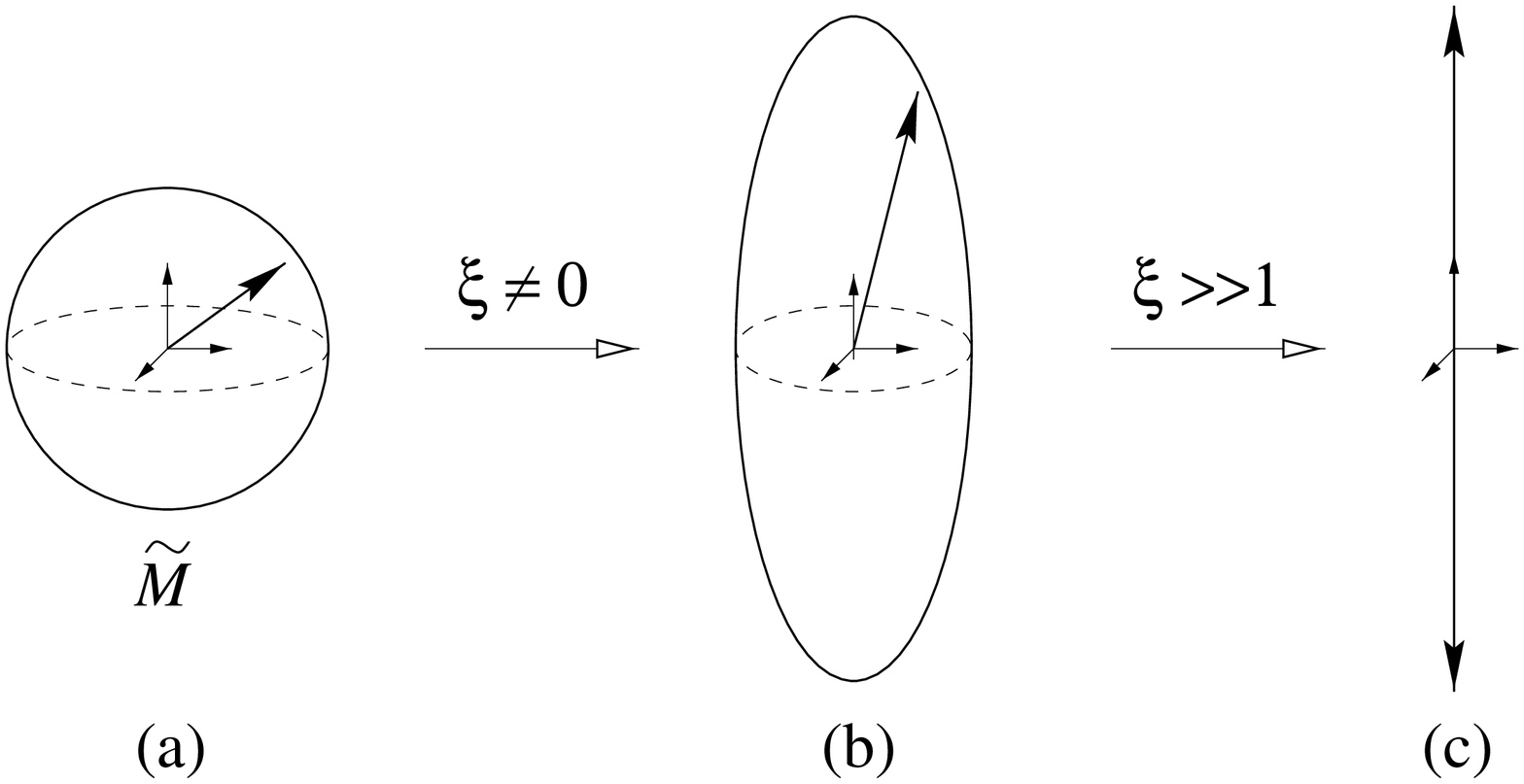,width=12cm,angle=0}%
         }

\vspace*{0.2in}

\newcaption{\footnotesize
 A schematic representation of the flavor moduli space 
 realized topologically as $S^2 \subset \R^3$, 
 showing the dependence of the induced metric on the hierarchical
 structure of the masses: (a) equal masses implying SO(3) isometry;
 (b) $\xi\neq 0$ implying SO(2) isometry; and (c) the decoupling
  limit where $\xi \gg 1$. The overall scale has been chosen for
 convenience. The bold arrows indicate an example of 
 the dominant flavor mode profile in the wall.}
\end{figure}
\noindent

In the process of decoupling the heavy flavor, $m_2\gg m_1$, 
the ellipticity parameter $\xi\sim \sqrt{m_2/m_1}$ diverges, and
the ellipsoidal metric on $S^2$ becomes singular. As shown
schematically in Fig.~3, the corresponding moduli are then ``frozen'' 
with two possible orientations, thus reproducing the expected result 
that there are two inequivalent 1-walls in this theory, and explaining
why a direct analysis of the SU(2) theory with 1 flavor would 
uncover two inequivalent solutions, with no additional moduli. This
latter result for the 1-flavor theory has been known for some 
time \cite{Stwo}. 

At this point it is worth contrasting the decoupling scenario with
our discussion in the previous section. Firstly, note that the
limit $m_2\gg m_1$ serves as a partial alternative to explicitly `lifting'
the additional flavor moduli by perturbing the theory with $N_f=N$ flavors, and
thus allows a direct calculation of the wall multiplicity. However, this 
freezing of moduli within weak coupling SQCD  is distinct from what 
may happen on integrating out all the matter fields, which we cannot
do here, and returning to the strong coupling regime in pure SYM theory. 
We see that despite the increasing ellipticity in this case, all mass scales
are small relative to $\La_2$, and the overall K\"ahler class of 
$\widetilde{\cal M}$ which scales as ${\cal O}(\La_2^2/\sqrt{m_1m_2})$
remains large. 

This picture of the hierarchical freezing of flavor modes, as some
subset of the matter fields are decoupled, allows us to make 
contact more generally with the picture of the wall
spectrum that emerges in SQCD with $N_f=N-1$ flavors. Recall
that in the hierarchical regime (\ref{mmatrix}), we could
simply integrate out the $N$-th flavor, leading to the
ADS superpotential (\ref{adsW2}) which exhibits an SU$(N-1)$ symmetry
if we set the remaining mass matrix proportional to the identity. In
this system, the wall multiplet structure is less explicit, but 
$\nu_k$ must necessarily be the same. As explained above, for SU(2)
the agreement follows straightforwardly from the fact that although
the additional flavor modes are frozen, this can happen in 
two possible ways, reproducing the two-wall spectrum obtained 
some time ago \cite{Stwo}.

The case of SU(3) gauge group broken via the Higgs mechanism through the 
introduction of two flavors can be understood in a similar manner.
This system been treated in some detail in the literature. There is
now a $2 \times 2$ meson matrix $M'$, and the flavor symmetry of 
the canonical K\"ahler potential is SU(2)$\times$SU(2) provided
the mass matrix $m'$ is chosen proportional to the identity. 
There are three vacuum states at $\langle M' \rangle 
\propto \omega_3^k\,\Id_2 $ ($k=0,1,2$), and this theory again
possesses only minimal 1-wall solutions.

In searching for classical BPS configurations, it is natural
to first introduce a diagonal {\em ansatz}, namely 
$(M')^g_f =M\delta^g_f$,  ($g,f=1,2$). Such field configurations will
not break the flavor SU(2), and are flavor-symmetric domain walls.
Consequently, there are no massless excitations on the wall
worldvolume, other than the translational modes. 
Numerical analysis in \cite{ASone,CM,CHMM,AStwo} demonstrated the existence of
a unique flavor-symmetric solution. However, this is not the end of
the story as the symmetric ansatz should be relaxed to find all
the possible 1-wall solutions. If one demands simply that
$(M')^g_f$ is diagonal, with $(M')^1_1 \neq (M')^2_2$, then 
additional solutions will arise in pairs by permutation of the
fields. Perturbative analysis indicates that there are 
at most four trajectories emanating from each vacuum, and an analysis
along the lines of \cite{hiv} demonstrates that only two of these
interpolate to the second vacuum providing true wall
configurations. This conclusion is backed up by explicit 
numerical solutions found\footnote{\,These authors work instead with a
Taylor-Veneziano-Yankielowicz superpotential, but one can compare the 
results for small $m'$.} in
\cite{CHMM}  which were confirmed in \cite{AStwo}. 
More generally, if we retain the dependence on the full $2 \times 2$
meson matrix, the flavor asymmetric ansatz will break the SU(2)
symmetry down to U(1), inducing flavor moduli
parametrizing a $\C$P$^1$ sigma model on the worldvolume. 
When regulated in the infrared, the Witten index is equal
to two, which is consistent with the findings above. 
Thus, in total there are three inequivalent solutions in agreement with
our earlier results. In accord with our discussion above regarding
freezing of the moduli associated with decoupled fields, 
we see here that the full moduli space $\C$P$^2$ of the
3-flavor theory is reduced to $\C$P$^1$ in the hierarchical mass
regime due to the reduction in flavor symmetry. Accounting correctly
for the frozen modes ensures that the result for the index in each
case is, of course, the same.

This counting of minimal walls, using an unconstrained parametrization of
the moduli in the $N-1$ flavor theory, is easily extended.
For gauge group SU$(N)$, there is an \mbox{$(N-1)\times(N-1)$}
meson matrix and an explicit SU$(N-1)$ flavor symmetry. Once again there is a 
unique flavor symmetric wall, while flavor asymmetric walls 
induce Goldstone modes parametrizing a $\C{\rm P}^{N-2}$ sigma model
via the broken flavor symmetry. The corresponding Witten index is
$N-1$, leading again to $1+(N-1)=N$ possible 1-wall configurations. Thus the
$N$-plet observed in Sec.~\ref{sec:four} decomposes in this case as
{\boldmath $N \longrightarrow 1+({N-1})\,$}.

\section{Discussion}
\label{sec:six}

Deforming SYM theory to SQCD in the Higgs phase
has allowed us to tune the symmetry structure so
that classically there was a moduli space of BPS domain walls,
enabling a robust calculation of the wall multiplicity given a
suitable infrared regulator. Note that this genuine weak coupling approach is 
in contrast to others for which relevant fields fail to remain weakly 
coupled throughout the wall trajectory. The results we obtained are 
consistent with those deduced by Acharya and Vafa \cite{av} using 
a string dual construction, and the worldvolume description has 
intriguing parallels with this work on
which we will elaborate further below. 

In particular, we will finish with a few comments on the dynamics 
of the translational moduli, an issue that we have suppressed thus far. 
Specifically, while the center of mass modulus certainly
decouples, one can also study the formation of composite 2-walls from
primary 1-walls with adjacent phase boundary conditions. We have 
emphasized that in this system the positions of the constituents
are not moduli (in contrast to certain \ntwo domain wall systems 
\cite{tong}), but one can still set up an unstable configuration 
\cite{manton} where the two constituents are well separated and observe the
interactions which will be sensitive to the SQCD spectrum.
Moreover, one can arbitrarily suppress the binding energy in the
large $N$ limit. The binding
energy per unit volume follows from the BPS formula,
\be
 T_2 - 2 T_1 = -\frac{3\pi \La^3}{4N} + {\cal O}(N^{-3}),
\ee
and thus the walls decouple at large $N$. If we set up such
an unstable configuration and allow the walls to evolve to a bound state,
there is a moduli space transition,
\be
 2(\R \times \C{\rm P}^{N-1}) \longrightarrow
  \R \times G(2,N), \label{bind}
\ee
which leads to a reduction in bosonic moduli. A remarkable feature of the
Grassmannian sigma model is that this reduction in moduli has a  
consistent interpretation in terms of an 
enhanced gauge symmetry for the composite. More precisely, if we 
formulate the corresponding gauged linear sigma models, the above 
transition corresponds directly to the gauge symmetry enhancement
U(1)$\times$U(1)$\,\longrightarrow\,$U(2), recalling that the nonlinear
model is realized in the limit in which the gauge kinetic term
decouples. In any regime where these gauge modes become dynamical, 
say at 1-loop, this picture becomes quite
consistent with the construction of \cite{av}, and indeed more
generally with any realization of $k$-walls in terms of 
$D$-branes \cite{witten97}. 

The leading behavior of the potential on the asymptotic moduli space of 
two 1-walls is calculable within the SQCD regime, and aspects of the
binding process in (\ref{bind}) can be studied within the 
context of the corresponding \none gauged linear sigma model 
\cite{wit93,wit94}, but we will defer discussion \cite{rsv2} of these features,
and other details of the worldvolume dynamics. 
It is important to note that, while this analysis
is tractable for small $m$ in the SQCD regime, the lack of 
supersymmetric constraints on the SQCD K\"ahler 
metric makes extrapolation to SYM theory at 
any more than a speculative level fraught with difficulty. This is 
why the worldvolume index has a privileged position as essentially 
the only protected quantity that we are guaranteed can survive 
this transition.

\bigskip
\centerline{\bf Acknowledgments}
\bigskip

We would like to thank B.~Acharya, J.~Gauntlett, F.~Goldhaber, A.~Gorsky, 
A.~Losev, C.~Nu\~nez, A.~Smilga, M.~Strassler, D.~Tong, C.~Vafa, 
T.~ter Veldhuis, and 
A.~Yung for many useful discussions and/or comments on the manuscript.
The work of M.S. and A.V. was supported in part by DOE grant DE-FG02-94ER408.

\end{document}